\documentclass[preprint]{iucr}              

\usepackage{amssymb}
\usepackage{amsmath}
\usepackage{amsthm}
\usepackage{siunitx}

\newcommand{\ldm}[1]{\textcolor{magenta}{#1}}

\usepackage{soul,color}

     \journalcode{J}              

\begin{document}                  



\title{Computing virtual dark-field X-ray microscopy images of complex discrete dislocation structures from large-scale molecular dynamics simulations}
\shorttitle{DFXM for DD structures}


\author[a,b]{Yifan}{Wang}
\author[c]{Nicolas}{Bertin}
\author[a,b]{Dayeeta}{Pal}
\author[a,b]{Kento}{Katagiri}
\author[a,b]{Sara J.}{Irvine}
\author[c]{Robert E.}{Rudd}
\cauthor[a,b]{Leora E.}{Dresselhaus-Marais}{leoradm@stanford.edu}

\aff[a]{Department of Materials Science and Engineering, Stanford University, 476 Lomita Mall, Stanford, CA, 94305, \country{USA}}
\aff[b]{SLAC National Accelerator Laboratory, Menlo Park, CA, 94025, \country{USA}}
\aff[c]{Lawrence Livermore National Laboratory, L-286, 7000 East Ave., Livermore, CA, 94550, \country{USA}}









\maketitle                        



\begin{abstract}
Dark-field X-ray Microscopy (DFXM) is a novel diffraction-based imaging technique that non-destructively maps the local deformation from crystalline defects in bulk materials.
While studies have demonstrated that DFXM can spatially map 3D defect geometries, it is still challenging to interpret DFXM images of the high dislocation density systems relevant to macroscopic crystal plasticity.
This work develops a scalable forward model to calculate virtual DFXM images for complex discrete dislocation (DD) structures obtained from atomistic simulations.
Our new DD-DFXM model integrates a non-singular formulation for calculating the local strain from the DD structures and an efficient geometrical optics algorithm for computing the DFXM image from the strain.
We apply the model to complex DD structures obtained from a large-scale molecular dynamics (MD) simulation of compressive loading on a single-crystal silicon.
Simulated DFXM images exhibit prominent feature contrast for dislocations between the multiple slip systems, demonstrating the DFXM's potential to resolve features from dislocation multiplication. 
%
The integrated DD-DFXM model provides a toolbox for DFXM experimental design and image interpretation in the context of bulk crystal plasticity for the breadth of measurements across shock plasticity and the broader materials science community.
\end{abstract}



\section{Introduction}
\label{sec:intro}

In crystalline materials, dislocation motion is the primary source of plastic deformation~\cite{hull_introduction_2011}.
The interaction and multiplication of dislocations lead to high dislocation density in the material, 
critical to many macroscopic crystal plasticity behaviors such as strain hardening~\cite{kocks_physics_2003}, creep~\cite{blum_dislocation_2009}, and shock plasticity~\cite{kuokkala_2_2024}.
%
%
Significant developments in computational techniques over the past few decades have enabled numerical simulations of dislocation behaviors in crystals that span multiple lengths and time scales.
Dislocation dynamics (DD) simulations are used as mesoscale proxies of the atomistic mechanisms detected by MD simulation and significantly push the limit of the time scale to $\qty{1}{\micro\meter}$,
but it is uncertain how many mechanisms from atomistic simulations are not captured and how important their effects are, such as jogs and point defects from dislocation interactions~\cite{bertin_frontiers_2020}.
On the other hand, atomistic simulation overcomes this uncertainty and captures all possible atomic mechanisms in response to external loading.
Large-scale MD simulation has demonstrated its ability to simulate dislocation evolution and interaction statistically representing crystal plasticity at a macroscopic length scale~\cite{zepeda-ruiz_probing_2017}.
Due to the great computational demand, long timescales 
are hardly accessible by MD simulations.
State-of-the-art atomistic simulation of strain-hardening can reach the lowest strain rate of $\qty{e7}{s^{-1}}$~\cite{zepeda-ruiz_atomistic_2021},
orders-of-magnitude higher than typical quasi-static mechanical tests of materials ($\qtyrange[range-units = single]{e-5}{e0}{s^{-1}}$),
but this strain rate falls in the range of high-velocity impact experiments involving shock-wave propagation and shock plasticity~\cite{meyers_dynamic_1994}.
The unique high-strain-rate condition makes it possible to directly compare the shock deformation experiments and MD simulations under similar conditions. 
However, there are still significant discrepancies in dislocation behaviors between MD simulations and experiments under shock plasticity conditions~\cite{meyers_chapter_2009},
due to limitations on the current experimental characterizations of dislocation behaviors.

Traditionally, \emph{ex-situ} direct observation of dislocations in metals is performed under transmission electron microscopy (TEM) post-mortem (after a shock)~\cite{cao_effect_2005}, 
%
%
fundamentally missing the capability to investigate transient dislocation mechanisms.
Although the limit of strain rate has recently been pushed to $\qtyrange[range-units = single]{e3}{e4}{s^{-1}}$ for \emph{in-situ} TEM~\cite{voisin_situ_2020},
capturing the timescale of MD deformation and its transient mechanisms ($<\qty{1}{ns}$) remains challenging for TEM detectors.
%
%
%
%
%
%
%
On the other hand, synchrotron-based ultrafast \emph{in-situ} X-ray diffraction has been the primary measurement to characterize lattice distortion of high-strain-rate deformation over the past decades~\cite{kalantar_developing_2000, wehrenberg_situ_2017},
%
and it has been routinely extended to the analysis of defect microstructures for crystal plasticity during shock~\cite{turneaure_twinning_2018, pascarelli_materials_2023},
because X-ray diffraction (XRD) is sensitive to the local strain fields (deformation) induced by defect (e.g., dislocation) arrangements~\cite{jakobsen_formation_2006, bertin_computation_2018}.
%
However, XRD only provides signals from the statistical average of the defect behavior within a large volume, introducing ambiguity in interpreting and identifying specific defect mechanisms because the spatial resolution is missing.
%

While classical XRD is not spatially resolved, the novel dark-field X-ray microscopy (DFXM) has emerged as a new full-field X-ray diffraction-based imaging technique.
By placing an X-ray objective lens along a diffraction peak, DFXM maps real-space images to 3D real-space microstructures in bulk crystals~\cite{poulsen_x-ray_2017}.
Applications of this method include mapping grain orientations in polycrystal material~\cite{simons_dark-field_2015},
and dislocation structures in a single crystal~\cite{jakobsen_mapping_2019, dresselhaus-marais_situ_2021, yildirim_extensive_2023}.
Recent development of DFXM used X-ray free electron lasers (XFEL) with high spatial ($<\qty{1}{\mu m}$) and temporal ($<\qty{100}{fs}$) resolutions~\cite{holstad_x-ray_2022, dresselhaus-marais_simultaneous_2023},
enabling the \emph{in-situ} direct observation of dislocation microstructures and dynamics during laser-driven acoustic waves onto a single-crystal diamond sample.
This approach offers a first step in technique development towards \emph{in-situ} shock DFXM experiments. 
To enable \emph{in-situ} shock imaging, we require theory insights into the contrast mechanisms of the high-dislocation-density regimes of the shock plasticity.
%
%
%
As such, DFXM has primarily been applied to low dislocation density structures,
%
%
such as the recrystallization process~\cite{lee_multiscale_2024}, 
%
high-temperature annealing~\cite{yildirim_extensive_2023}, 
%
and the incipient plastic deformation under low strain-rate loading~\cite{zelenika_3d_2024}, 
%
DFXM has been used to quantify dislocation densities for high-dislocation-density systems to investigate the stress concentration and strain localization due to intragranular microstructural features~\cite{gustafson_quantifying_2020, gustafson_revealing_2023}. 
%
However, dislocation structures within a grain are highly heterogeneous.
Although the cutting-edge spatial resolution can reach a value below $\qty{100}{nm}$ such that many low-dislocation density systems can be investigated, identifying dislocation structures and mechanisms from DFXM images of high-dislocation density systems remains unclear.
%
%
%
To quantitatively connect the contrast in DFXM images to individual dislocation segments in such system based on their statistical elastic distortion field, 
%
it is necessary to develop a forward model of the DFXM image from complex DD structures based on physical simulations of dislocation evolution relevant to crystal plasticity during high-strain-rate shock compression,
such as large-scale MD simulations~\cite{zepeda-ruiz_atomistic_2021, luu_shock-induced_2020, holian_plasticity_1998} 
and coarse-grained discrete dislocation dynamics (DDD) simulations~\cite{shehadeh_multiscale_2005, sills_dislocation_2018, rao_large-scale_2019}.
%

In this work, we develop a scalable numerical model, named the non-singular DD-DFXM model, to compute virtual DFXM images of complex discrete dislocation (DD) structures from MD simulations.
Our new model integrates the geometrical-optics formulation for computing DFXM image from the deformation gradient field~\cite{poulsen_geometrical-optics_2021},
with the non-singular formulation for calculating the deformation of the DD structures~\cite{bertin_computation_2018}.
To demonstrate our model, we perform a large-scale MD simulation of high-strain-rate compression in silicon to simulate dislocation evolution under shock plasticity.
%
%
The DD networks are obtained from the saved atomic configurations of the MD deformation at different loading stages with increasing dislocation densities, 
and the DFXM images are simulated for two DD structures at low and high dislocation densities.
%
%
%
Our model provides a powerful toolbox to connect computational DD models to \emph{in-situ} DFXM imaging of dislocation behaviors relevant to crystal plasticity at high-strain rate deformation.

\section{Simulation and modeling approach}
\label{sec:method}

Poulsen \emph{et al.} (2021)~\cite{poulsen_geometrical-optics_2021} presented a geometrical optics formulation to translate the local deformation gradient field for a single dislocation inside a sample to the contrast mechanism produced by DFXM.
The geometrical optics model is numerically more efficient than the Takagi-Taupin equation-based wave-propagation method~\cite{carlsen_simulating_2022}, although it does not account for the effects of coherence and dynamical diffraction.
Recent work by Borgi, \emph{et al.} has demonstrated that the dynamical diffraction effects are not detrimental to visualize dislocations and that the kinematical approximations can be well justified in most cases~\cite{borgi_simulations_2024}.
So far, the DFXM community has focused on developing the optical formalism while using simple analytical expressions of the displacement vector field $\vec{u}$ or deformation gradient tensor field $\mathbf{F}^g$ of infinite straight dislocations~\cite{poulsen_geometrical-optics_2021}.
While these are representative of the effects of isolated defects, they can only construct limited numbers of geometries, such as arrays of dislocations~\cite{borgi_simulations_2024} and planar stacking faults~\cite{carlsen_simulating_2022}.
These expressions are based on continuous elasticity solutions with singularities at the dislocation core that introduce numerical artifacts, and it is challenging to develop complex dislocation structures based on these analytical solutions.
Therefore, to overcome these limitations, we apply a general non-singular formulation of the deformation gradient field for complex dislocation structures to the DFXM forward model.

Our numerical model integrates the established geometrical optics model~\cite{poulsen_geometrical-optics_2021} with the discrete dislocation (DD) model adapted from the convention of discrete dislocation dynamics (DDD) simulations.
In the DD model, the continuous curved dislocation lines are represented as discretized line segments with their Burgers vectors,
enabling the representation of a highly complex dislocation network and retaining the key microstructural features of the network~\cite{bertin_frontiers_2020}.
The DD model can be obtained from DDD simulation, or extracted directly from MD simulations~\cite{stukowski_automated_2012}.
The general continuum non-singular formulation for the DD model has been developed for over twenty years~\cite{cai_non-singular_2006}
%
%
and has been used to calculate virtual X-ray diffraction patterns~\cite{bertin_computation_2018}.
This section introduces the key formulations and the workflow of our integrated DD-DFXM forward model.
%

\subsection{Geometrical optics formulation of DFXM}
\label{sec:geometrical_optics}

\begin{figure}
     \centering
     \includegraphics[width=0.95\linewidth]{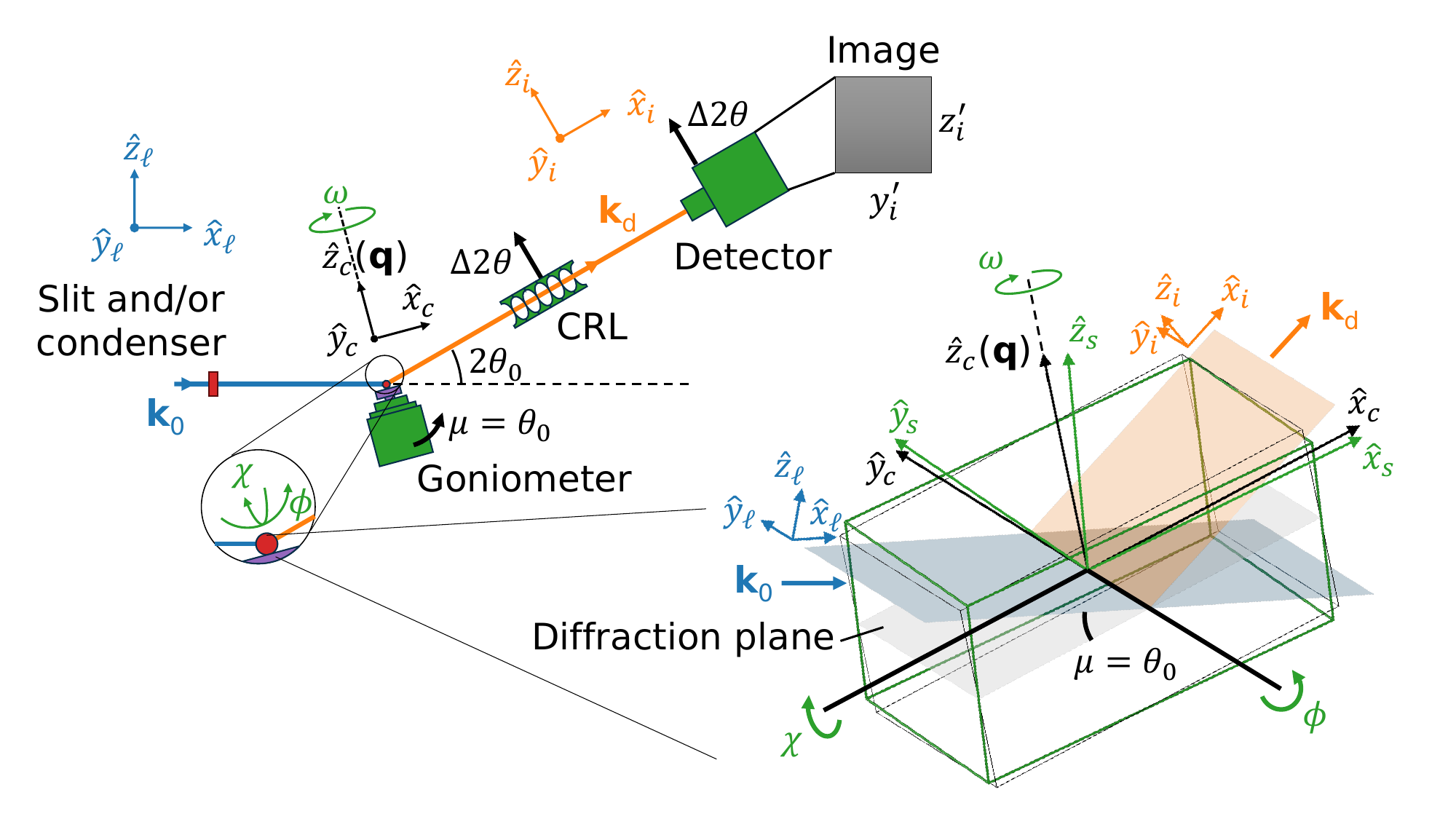}
     \caption{
     {\bf Geometry and coordinate systems of DFXM.}
     The laboratory coordinate system ($\hat{x}_\ell$, $\hat{y}_\ell$, $\hat{z}_\ell$) aligns with the incident beam ($\mathbf{k}_0$).
     While the imaging coordinate system ($\hat{x}_i$, $\hat{y}_i$, $\hat{z}_i$) diffracted beam ($\mathbf{k}_d$).
     The goniometer's pivot point and the sample are centered at the intersection of the incident ($\mathbf{k}_0$) and diffracted ($\mathbf{k}_d$) beams.
     The compound refractive lens (CRL) stack is the objective lens between the sample and the detector.
     The zoom-in figure presents the definition of different coordinate systems in the DFXM model explained in the main text.
     %
     %
     %
     %
     %
     %
     The figure is adapted from~\cite{poulsen_x-ray_2017}.
     }
     \label{fig:geometry}
\end{figure}

Figure~\ref{fig:geometry} illustrates the geometry of the DFXM optics.
The incident X-ray beam ($\mathbf{k}_0$) aligns with the $\hat{x}_\ell$-direction in the laboratory ($_{\ell}$) coordinate system $(\hat{x}_\ell, \hat{y}_\ell, \hat{z}_\ell)$,
while the diffracted beam ($\mathbf{k}_d$) aligns with the imaging coordinate system ($_i$).
%
%
The diffraction vector $\mathbf{q}$ is normal to the selected diffraction plane in the single crystal sample, aligning with the crystal coordinate system ($\hat{x}_c$, $\hat{y}_c$, $\hat{z}_c$).
%
%
The goniometer is associated with a base tilt $\mu$, an $\omega$ rotation around $\mathbf{q}$ and two tilts, $\phi$ and $\chi$.
The base-tilt of the goniometer $\mu=\theta_0$ affixes the sample at the selected specific diffraction peak, such that the crystal diffracts in the vertical diffraction conditions at an angle of $2\theta_0$.
Changing $\mathbf{q}$ requires remounting of the sample and adjustment of the pre-tilt angle $\mu$, while changing the tilts ($\phi, \chi$) and rotation ($\omega$) does not.
The sample coordinate system ($\hat{x}_s$, $\hat{y}_s$, $\hat{z}_s$) is defined to stick to the sample's surface normal vector.
Initially overlapping with the crystal coordinate system, the sample coordinate system rotates and tilts together with ($\phi, \chi, \omega$).
Through the objective lens stack placed along the diffraction beam direction, 
the coordinate of the intensity on the detector $I(y'_i, z'_i)$ is a magnification of the real-space coordinate $(y_i, z_i)$ in the imaging system.

The governing equation of the pixel intensity on the detector $I(y'_i, z'_i)$ can be calculated by integrating the real-space points $\mathbf{r}$, and the reciprocal space components for defining diffraction at those points, $\mathbf{q}$, via a 6D integral,
\begin{equation}
    I(y'_i, z'_i) = \iiint_{\mathbf{r}}\iiint_{\mathbf{q}}
        \Phi_d(\mathbf{r}, \mathbf{q})\,
        {\rm Res}(\mathbf{r}, \mathbf{q}, y'_i, z'_i)
        {\rm d}^3\mathbf{r}{\rm d}^3\mathbf{q},
\end{equation}
where $\Phi_d$ is the flux diffracting from a real-space position $\mathbf{r} = (x_i, y_i, z_i)$ (in the imaging coordinate system).
${\rm Res}(\mathbf{r}, \mathbf{q}, y'_i, z'_i)$ is the resolution function of the instrument and the material,
modeled as the fraction (a value between $[0,1]$) of the incident flux being diffracted and projected onto the detector.
The terms ${\rm d}^3\,\mathbf{r}{\rm d}^3\mathbf{q}$ describe an infinitesimal volume in the 6D real ($\mathbf{r}$) and reciprocal ($\mathbf{q}$) spaces.

\begin{figure}
     \centering
     \includegraphics[width=0.95\linewidth]{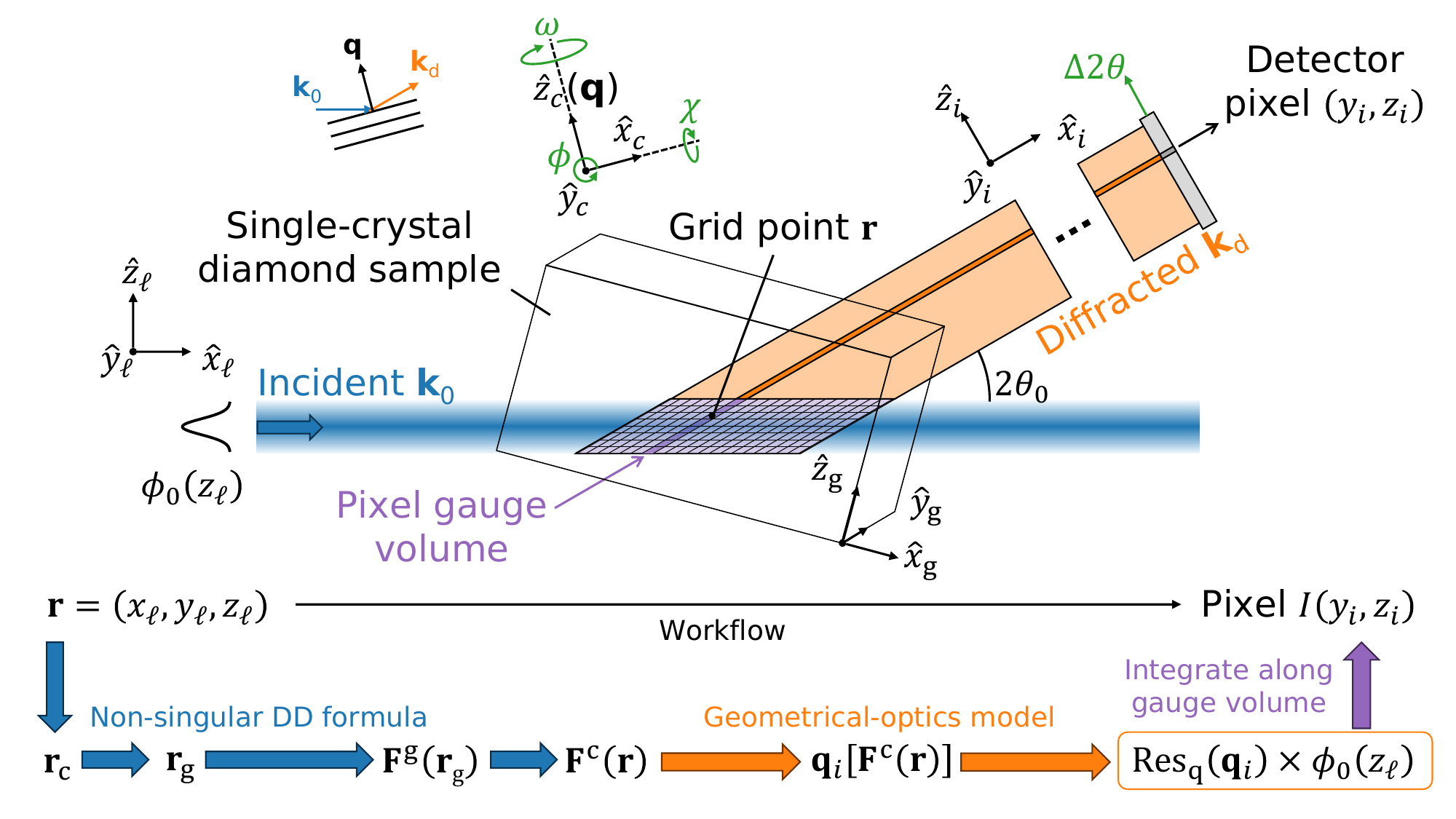}
     \caption{{\bf Schematic of the pixel integration \& workflow of the geometrical optics model.}
         The illuminated volume (light purple) is bounded by the incident beam width $L_z=6z_{\ell,{\rm rms}}$ and the field of view of the imaging system.
         %
         %
     }
     \label{fig:definition}
\end{figure}

The geometrical optics forward model of DFXM assumes kinematic diffraction and a continuous function for the deformation gradient field $\mathbf{F}^g(\mathbf{r})$ at the far-field.
Numerically, the gauge volume is discretized as small voxels according to the pixel size, as shown in Fig.~\ref{fig:definition}.
For each voxel, the center $\mathbf{r}$ represents its position, and a uniform deformation tensor value $\mathbf{F}^{\rm g}(\mathbf{r})$ through out the voxel volume represents the average deformation.
%
%
%
%
As a result of the kinematic diffraction assumption, the diffracted wave vector of each voxel $\mathbf{q}[\mathbf{F}^{\rm g}(\mathbf{r})]$ due to the deformation represents only one single point in the reciprocal space.
The full 6D resolution function can also be decomposed into two 3D resolution functions: the spatial ${\rm Res}_{\mathbf{r}}$ and the reciprocal ${\rm Res}_{\mathbf{q}}$ components.
Finally, a simplified, low-dimensional expression for the intensity can be written as (Eq.~(58) in \cite{poulsen_geometrical-optics_2021}),
\begin{align}
    I(y'_i, z'_i) = C &
        \int_{y_i}\int_{z_i} {\rm PSF}(y_i - y'_i, z_i - z'_i) \nonumber \\
    &   \int_{x_i} \Phi_0(\mathbf{r}_i) \rho(\mathbf{r}_i)
            {\rm Res}_{\mathbf{q}_i} \left[\mathbf{q}_i(\mathbf{r}_i) + 
            \Delta \mathbf{q}_i(y'_i, z'_i)\right]
            {\rm d}x_i{\rm d}z_i{\rm d}y_i
    \label{eqn:forward}
\end{align}
where the constant pre-factor $C$ represents the scattering terms such as the Lorentz and polarization factors.
$\Phi_0(\mathbf{r}_i)$ is the incident flux at position $\mathbf{r}_i$, and $\rho(\mathbf{r}_i)$ is the local normalized density of the material at that position.
The density term $\rho(\mathbf{r}_i)$ is important for shock compression as that often experiences $\unit{GPa}$ pressures ($>\qty{100}{GPa}$ for diamond)~\cite{mcwilliams_strength_2010}.
The change of material density due to a shock loading wave can be treated separately and incorporated by leaving the spatial $\rho(\mathbf{r}_i)$ term in the equation~\cite{wood_molecular_2017}.
For simplicity, in this work, we assume the solid material's density $\rho(\mathbf{r}_i)=\rho_0$ is a constant in the simulation cell.

The real-space resolution function ${\rm Res}_{\mathbf{r}}$ is re-written in the image coordinate system as a two-dimensional (2D) point spread function, ${\rm PSF}(y-y',z-z')$,
representing the spread of the projected intensity of the diffracted beam to the detector.
%
%
%
The PSF can be further simplified as a 2D delta function, i.e., ${\rm PSF}(y-y',z-z')=C'\delta(y-y')\delta(z-z')$,
so that the diffraction can be simplified as an 1D integration along the $\hat{x}_i$ direction~\cite{poulsen_geometrical-optics_2021}.
Thus, we will use $(y_i, z_i)$ and $(y'_i,z'_i)$ interchangeably to represent a pixel in the following text.
%
%
%
%
%
%
%
The 3D reciprocal-space resolution function ${\rm Res}_{\mathbf{q}}(\mathbf{q}_i)$ represents the spread of the diffraction peak in the reciprocal space according to DFXM settings,
which can be numerically determined using a Monte-Carlo ray-tracing algorithm~\cite{poulsen_geometrical-optics_2021}, see Appendix~\ref{sec:ray_tracing}.
The ray-tracing algorithm for reciprocal-space resolution function ${\rm Res}_{\mathbf{q}}$ only needs to be calculated once for a given DFXM setting, such as X-ray beam, apertures, lenses, and the crystal diffraction angle.
The pre-calculated ${\rm Res}_{\mathbf{q}}$ does not need to be re-calculated for goniometer tilts ($\phi,\chi,\omega$) or sample deformation, as long as the deformation does not induce additional rotation to the diffraction plane.

The reciprocal space wavevector (in the image coordinate) $\mathbf{q}_i(\mathbf{r}_i)$ is a function of the real-space vector $\mathbf{r}_i$,
determined by the local deformation gradient field $\mathbf{F}^s$, the rotation of the goniometer $(\phi, \chi)$ ($\omega = 0$ in our simplified geometry), and the motion of the detector and lens $\Delta\theta$, as shown in Fig.~\ref{fig:geometry}.

Departing from \cite{poulsen_geometrical-optics_2021}, we re-define a few coordinate systems for the concepts specific to this work.
The grain system ($^g$) is defined as the coordinate system native to the MD simulations, aligned with the identity matrix $\mathbf{I}=[\hat{x}_{\rm g}, \hat{y}_{\rm g}, \hat{z}_{\rm g}]$ in the Miller's indices ($\hat{x}_{\rm g}=[100]$, $\hat{y}_{\rm g}=[010]$, $\hat{z}_{\rm g}=[001]$).
The sample coordinate system ($^s$) is defined such that the $\hat{z}_s$-direction always aligns with the normal vector of the sample surface, initially overlapping with the crystal system ($^c$), but rotating and tilting together with the goniometer $(\phi,\chi,\omega)$.
%
The deformation gradient tensor in the grain coordinate system $\mathbf{F}^{\rm g}$ is defined as,
\begin{equation}
    \mathbf{F}^{\rm g} = \nabla\mathbf{u} + \mathbf{I}
\end{equation}
To convert the deformation gradient tensor between the sample and grain coordinate systems requires a rotation matrix $\mathbf{U} = [\hat{x}_s, \hat{y}_s, \hat{z}_s]^T$, 
where $\hat{x}_s$ and $\hat{y}_s$ are the sample mounting direction (normalized) written in the grain coordinate system (Miller indices).
Thus, the vectorial and tensorial conversion between the two coordinate systems can be written as,
\begin{align}
    \mathbf{r}_s &= \mathbf{U}\mathbf{r}_g \\
    \mathbf{F}^s &= \mathbf{UF}^g\mathbf{U}^T
\end{align}
The micro-mechanical model we use in this work to calculate the $\mathbf{F}^{\rm g}$ given a discrete dislocation structure is discussed in Section~\ref{sec:non_singular}.

In this work, the tilt angles ($\phi$, $\chi$, and $\Delta\theta$) are relatively small ($\sim\qty{e-4}{rad}$).
On the one hand, the deformation gradient tensor, $\mathbf{F}$, represented in the sample ($^s$, the green coordinates in Fig.~\ref{fig:geometry}) and the crystal ($^c$, the black coordinates in Fig.~\ref{fig:geometry}) coordinate systems are approximately the same, i.e., $\mathbf{F}^s\approx\mathbf{F}^c$.
In the following text, we will use $\mathbf{F}^c$ and $\mathbf{F}^s$ interchangeably, as their conversion has minimal impact on the signals of interest to this work. We note that this simplification speeds up the run-time of these methods significantly, despite adding relatively little change to the signal.
We instead account for these tilt angles as a first-order correction to the diffracted wavevector $\mathbf{q}_c$ in the crystal system~\cite{poulsen_geometrical-optics_2021},
\begin{align}
     \mathbf{q}_i&=\mathbf{R}(\theta_0)\cdot\mathbf{q}_c
             = \mathbf{R}(\theta_0)\left[
                    \mathbf{q}_s + \begin{pmatrix}
                        \phi - \Delta\theta \\ \chi \\ \Delta\theta\cot(\theta_0)
                    \end{pmatrix}
               \right]  \nonumber \\
            &= \mathbf{R}(\theta_0)\left[
                    \mathbf{U}\left([\mathbf{F}^g]^{-T}-\mathbf{I}\right)\mathbf{Q}_{hkl} + \begin{pmatrix}
                        \phi - \Delta\theta \\ \chi \\ \Delta\theta\cot(\theta_0)
                    \end{pmatrix}
               \right]
    \label{eqn:diff_wave_vec_i}
\end{align}
where the rotation matrix from the reference diffraction direction ($\mathbf{q}_c$) to the imaging direction ($\hat{x}_i$) is,
\begin{equation}
    \mathbf{R}(\theta_0) = \begin{bmatrix}
        \cos\theta_0 & 0 & \sin\theta_0 \\
        0 & 1 & 0 \\
        -\sin\theta_0 & 0 & \cos\theta_0
    \end{bmatrix}
\end{equation}
and $\theta_0$ is the diffraction angle for the plane $(hkl)$, i.e., $\lambda = 2d_{hkl}\sin\theta_0$ for the undeformed lattice.
For diamond, the specific diffraction angles used in this work are $\theta_0(d_{004}) = \ang{24.08}$ and $\theta_0(d_{111}) = \ang{10.03}$, assuming photon energy of $\qty{17.29}{eV}$, all the other instrumental parameters are consistent with the setup in~\cite{dresselhaus-marais_situ_2021, poulsen_geometrical-optics_2021}.
The corresponding rotation matrix $\mathbf{U}$ for the two diffraction planes in this work are:
\begin{equation}
     \mathbf{U}_{004} = \begin{bmatrix}
          1 & 0 & 0 \\
          0 & 1 & 0 \\
          0 & 0 & 1
     \end{bmatrix},\quad
     \mathbf{U}_{111} = \begin{bmatrix}
          1/\sqrt{2} & -1/\sqrt{2} & 0 \\
          1/\sqrt{6} & 1/\sqrt{6} & -2/\sqrt{6} \\
          1/\sqrt{3} & 1/\sqrt{3} & 1/\sqrt{3}
     \end{bmatrix}
\end{equation}
Note that the deformation gradient tensor $\mathbf{F}^g$ only needs to be evaluated once for the gauge volume with zero tilts $\phi=\chi=2\Delta\theta=0$,
since these tilts have been accounted for as a first-order approximation in Eq.~(\ref{eqn:diff_wave_vec_i}).
Therefore, the simplified forward model equation can be written as a 1D integral along the ${x}_i$-direction for calculating the pixel intensity at point $(y_i,z_i)$ on the detector,
\begin{align}
    I_{\rm fwd}(y_i, z_i) &\equiv \frac{I(y_i, z_i)}{CC'\rho_0} \nonumber \\
    &= 
    \int_{x_i}\Phi_0(\mathbf{r}_i)
            {\rm Res}_{\mathbf{q}_i} (\mathbf{q}_i\left[\mathbf{F}^g(\mathbf{r}_i); (\phi,\chi,\omega,\Delta\theta)\right])
            {\rm d}x_i,
    \label{eqn:dfxm_forward_final}
\end{align}

The schematics in Fig.~\ref{fig:definition} demonstrate the process simulated by the forward model of DFXM.
The incident beam $\mathbf{k}_0$ is modeled as a Gaussian intensity $\Phi_0(z_\ell)$ in the $\hat{z}_\ell$ direction with a full-width at a half max of ${\rm FWHM}_z=\qty{600}{nm}$, equivalent to a variance (root mean square) of $z_{\ell, {\rm rms}}={\rm FHWM}_z/2.35=\qty{255.3}{nm}$.
The intersection between the incident beam (the Gaussian width $L_z=6z_{\ell,{\rm rms}}$, blue) and the volume bounded by the detector active area (light orange) in $\hat{y}_i$ and $\hat{z}_i$ directions (i.e., the field of view of the imaging system) defines the total gauge volume (light purple).
Similarly, each pixel $(y_i, z_i)$ in the imaging coordinate system has an associated gauge volume that is the intersection (dark purple) of the incident beam width (blue) and the volume bounded by the active area of that pixel (dark orange).
The pixel intensity $I_{\rm fwd}(y_i, z_i)$ is then calculated by integrating the diffracted intensity over the pixel gauge volume, as defined by the spatial extent of the pixel $(y_i, z_i)$. 
%
%
This integration is done by discretizing the pixel gauge volume into an array of voxels (white box in Fig.~\ref{fig:definition}).
Each discretized voxel is labeled as a discretized field point $\mathbf{r}$, which can be equivalently represented in both the crystal $\mathbf{r}_c=(x_c,y_c,z_c)$ and the imaging $\mathbf{r}_i=(x_i,y_i,z_i)$ coordinate systems.
We note that while the crystal system is most convenient to define the micromechanical model, the imaging system natively offers a unique axis of integration along $\hat{x}_i$ for the pixel integration.
The average deformation tensor of each voxel $\mathbf{F}^{\rm g}(\mathbf{r})$ is then needed as an input to the geometrical model.
In the next section, we will discuss the formulation of $\mathbf{F}^{\rm g}(\mathbf{r})$ due to the DD structure in the simulation cell.
%
%
%

\subsection{Non-singular formulation of discrete dislocation structures}
\label{sec:non_singular}

\begin{figure}
     \centering
     \includegraphics[width=0.95\linewidth]{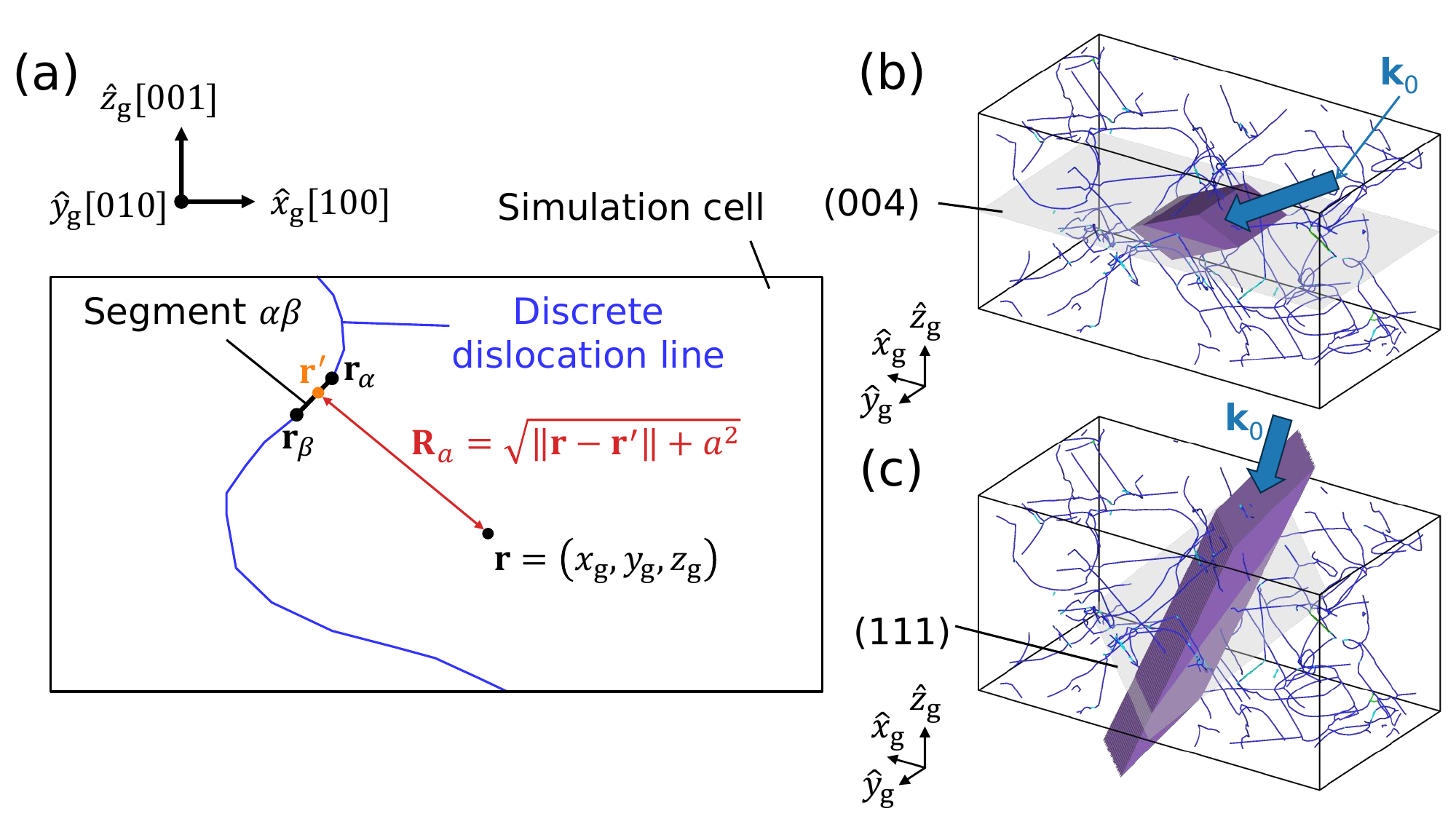}
     \caption{
         {\bf Schematics of the non-singular formulation of DD model.}
         (a) The simulation cell is drawn in the grain coordinate system ($^{\rm g}$),
         and the 3D curved dislocation lines in the simulation box are discretized as small line segments.
         %
         %
         (b) The purple gauge volume for the diffraction plane $(004)$ is illustrated in a simulated DD structure.
         (c) The gauge volume for the diffraction plane $(111)$.
         The blue arrow indicates the direction of the incident beam $\mathbf{k}_0$, and the gray plane represents the diffraction plane.
     }
     \label{fig:non_singular_disl}
\end{figure}

%
%
%
This section introduces the non-singular deformation gradient formulation for the discrete dislocation (DD) model used in DDD simulations.
In DD model, the continuous 3D dislocation curves are discretized into dislocation line segments.
Each segment $\alpha\beta$ consists of the position of the starting point $\mathbf{r}_\alpha$, the end point $\mathbf{r}_\beta$, 
and the Burgers vector ($\mathbf{b}$) following the left-hand start-finish (LH/SF) convention~\cite{anderson_theory_2017, stukowski_automated_2012}, as shown in Fig.~\ref{fig:non_singular_disl}.
The line-sense vector that defines dislocation segment direction is defined as $\hat{\xi} = (\mathbf{r}_{\alpha} - \mathbf{r}_{\beta})/\|\mathbf{r}_{\alpha} - \mathbf{r}_{\beta}\|$.
%
%

The total deformation gradient $\mathbf{F}^g$ of a field point $\mathbf{r}$ (in the grain coordinate system) can be calculated as the superposition of the contribution from all the segments $\alpha\beta$ in the discrete dislocation structure,
\begin{equation}
    \mathbf{F}^g(\mathbf{r}_g) = \nabla\mathbf{u}(\mathbf{r}_g) + \mathbf{I} = \sum_{\alpha\beta}\nabla\mathbf{u}^{\alpha\beta}(\mathbf{r}_g) + \mathbf{I}
    \label{eqn:deformation_tensor}
\end{equation}

The continuous elasticity solution of dislocations that has been used previously for DFXM simulations ignores the dislocation core size and introduces numerical artifacts (singularity) near the dislocation core.
This can destabilize the numerical computation, necessitating that our dense dislocation networks require an alternate approach when translating our DD model to the continuum scale.
Bertin \& Cai~\cite{bertin_computation_2018} derived a non-singular formulation for an arbitrary dislocation network to overcome this issue.
The displacement gradient (in Einstein's notation, see Appendix~\ref{sec:einstein}) due to the dislocation segment $\alpha\beta$ can be written at a spatial point $\mathbf{r}$ as,
\begin{align}
    u_{m,l}^{\alpha\beta}(\mathbf{r}_g) = 
        &- \frac{1}{8\pi}\int_{\mathbf{r}_\alpha}^{\mathbf{r}_\beta} b_m \epsilon_{jlk} R_{a,ppj}\,{\rm d}r'_k 
         - \frac{1}{8\pi}\int_{\mathbf{r}_\alpha}^{\mathbf{r}_\beta} b_i\epsilon_{mik} R_{a,ppl}\,dr'_k \nonumber\\
        &- \frac{1}{8\pi(1-\nu)}\int_{\mathbf{r}_\alpha}^{\mathbf{r}_\beta} b_i\epsilon_{ijk}R_{a,mjl}\,dr'_k
    \label{eqn:dispgrad_seg}
\end{align}
where $\mathbf{r}_\alpha$ and $\mathbf{r}_\beta$ are the positions defining the endpoints of the dislocation segment $\alpha\beta$,
and we choose the average Poisson's ratio $\nu = 0.2$ for diamond~\cite{klein_anisotropy_1992}.
$R_a$ is the non-singular distance between the spatial point $\mathbf{r}$ and the point $\mathbf{r}'$ on the integral path (segment $\alpha\beta$), defined as,
\begin{equation}
    R_a = \sqrt{R^2 + a^2} = \sqrt{\|\mathbf{r}-\mathbf{r}'\|^2 + a^2}
\end{equation}
where $a$ is a parameter representing the radius of the dislocation core.
The core size, $a$, defines the region that is not well described by the elastic solution, and is typically selected as the magnitude of Burgers vector $b_{\rm mag}$ for most crystalline materials~\cite{cai_non-singular_2006}.
For crystals with dissociated partial dislocations such as face-centered cubic (FCC),
the $a=b_{\rm mag}$ still applies, and any partial dislocations (formed when sufficiently low stacking-fault energies are present) are considered as individual dislocation lines.
%
%

In MD simulations, the boundaries of the simulation box are often set to satisfy periodic boundary conditions (PBC), where an object moving out of one side of the box will re-enter the box from the opposite side, to avoid surface effects and simulate the bulk material~\cite{bulatov_computer_2006}.
We implement the PBC in our simulations so that the gauge volume can still capture the dislocation segments on the other side, even if extended outside the simulation box.
%
%

%
The detailed process for evaluating $I_{\rm fwd}(y_i,z_i)$ of a single pixel $(y_i,z_i)$ on the detector with Eq.~(\ref{eqn:dfxm_forward_final}) takes the following approach, as shown schematically in Fig.~\ref{fig:definition}:
\begin{enumerate}
    \item According to the beam width ($z_{\ell,{\rm rms}}$),
    the size $p_{\rm size}$ and the position $(y_i,z_i)$ of the pixel,
    determine the pixel gauge volume as an array of voxels $\mathbf{r}$ in the lab ($\mathbf{r}_\ell$), the crystal ($\mathbf{r}_c$), and the imaging ($\mathbf{r}_i$) coordinate systems,
    as shown in Fig.~\ref{fig:definition}.
    \item For each voxel $\mathbf{r}$,
    calculate the normalized incident beam flux $\Phi_0(\mathbf{r})$
    based on where it is located on the Gaussian distribution $\mathcal{N}(0,z_{\ell,{\rm rms}})$.
    \item Evaluate the displacement gradient $\nabla\mathbf{u}^{\alpha\beta}(\mathbf{r})$ due to every dislocation segment $\alpha\beta$ in the sample using Eq.~(\ref{eqn:dispgrad_seg}).
    \item Evaluate the local deformation gradient tensor $\mathbf{F}^g(\mathbf{r})$ at each voxel with Eq.~(\ref{eqn:deformation_tensor}) by summing the contributions from each segment $\alpha\beta$.
    \item \label{item:step_qi} Calculate the diffraction vector $\mathbf{q}_i$ for each voxel based on $\mathbf{F}^g$ and the goniometer tilts ($\phi$, $\chi$, $\Delta\theta$) using Eq.~(\ref{eqn:diff_wave_vec_i}).
    \item Determine the fraction of the beam incident intensity being diffracted to the detector using the pre-calculated resolution function ${\rm Res}_{\mathbf{q}_i}(\mathbf{q}_i)$.
    \item Integrate (i.e., sum) the diffracted beam intensity ($\Phi_0{\rm Res}_{\mathbf{q}_i}$) for all the voxels describing the pixel's gauge volume of $(\Delta y'_i, \Delta z'_i)$ to obtain the intensity on the detector using Eq.~(\ref{eqn:dfxm_forward_final}).
\end{enumerate}
The simulated 2D DFXM image is then compiled with the intensity of all the pixels $I_{\rm fwd}(y_i,z_i)$ in the imaging coordinate system.
%
%
The geometrical DFXM formalism in this and previous works is numerically efficient because its resolution function ${\rm Res}_{\mathbf{q}_i}$ and deformation gradient tensor $\mathbf{F}^g$ can be pre-calculated.
%
%
When the sample has deformed or the gauge volume has rotated sufficiently (changing $\omega$ or $\mu$), $\mathbf{F}^{\rm g}$ must be re-evaluated.
When there is any change to the optical parameters, ${\rm Res}_{\mathbf{q}_i}$ must be re-evaluated, such as changing the diffraction plane ($\theta_0$), the numerical apertures of the slit/condenser/CRL, or the energy/bandwidth/coherence of the X-ray beam.

\section{Results}
\label{sec:results}

In this section, we first test our integrated non-singular DD-DFXM algorithm on the test case of a triangular prismatic dislocation loop.
We then simulate and analyze the dislocation structures from the MD diamond-shock simulation to obtain the dislocation density and slip systems.
The DD-DFXM model is then applied to the dislocation structures from our material simulations.
We apply the rocking and rolling tilts of the goniometer to find good weak beam conditions for observing different types of dislocations.
Finally, we isolate dislocation features such as jogs and junctions and show they have recognizable features on the simulated DFXM image.

\subsection{Triangular prismatic dislocation loop}
\label{sec:triangular_loop}

\begin{figure}
     \centering
     \includegraphics[width=0.95\linewidth]{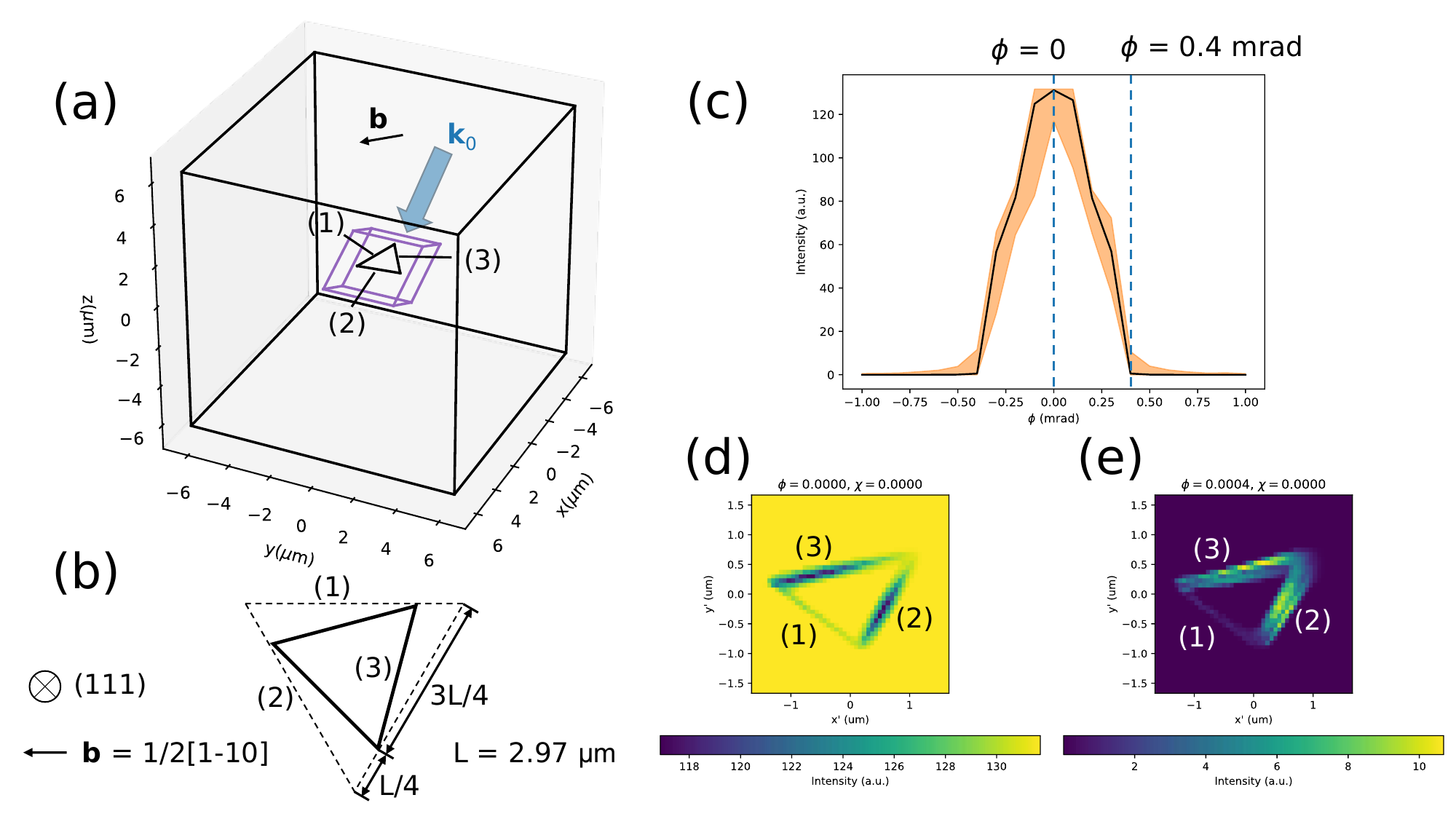}
     \caption{
     {\bf Test case of a prismatic triangular dislocation loop.}
     (a) 3D geometry of the test case, in which a triangular dislocation loop is located at the center of a cubic simulation box (size $\qty{12.61}{um}$).
     The gauge volume considered from the incident beam $\mathbf{k}_0$ is illustrated as the purple box.
     %
     %
     The inlet (b) shows the triangular loop on the $(111)$ slip plane,
     %
     %
     The dislocation loop is a slightly rotated equilateral triangle on the $(111)$ plane, illustrated as the solid black line with a Burgers vector of $\frac{1}{2}[1\bar{1}0]$.
     (c) Rocking curve of the DFXM images plotted as the average image intensity (black line), and an orange area indicating the min and max pixel intensity as functions of the $\phi$ tilt ($\chi = 0$).
     The strong ($\phi=0$) and the weak ($\phi=\qty{0.4}{\milli\radian}$) beam conditions are marked as dashed vertical lines.
     (d) Simulated DFXM image at the strong beam condition, 
     and (e) at the weak beam condition. 
     }
     \label{fig:triloop}
\end{figure}

To test and validate the numerical accuracy and efficiency of our integrated DFXM-DDD model, we begin with a test case of a triangular prismatic dislocation loop similar to the one discussed in~\cite{bertin_computation_2018}.
We consider an equilateral triangular dislocation loop on the $(111)$ plane, as shown in Fig.~\ref{fig:triloop}a-b.
The center of the triangle is on the origin $(0,0,0)$ of the simulation box with the edge length of $\qty{12.61}{\mu m}$ or $\num{50000}b$, where $b=\qty{0.2522}{nm}$ is the Burgers vector magnitude of diamond crystal.
The three segments are labeled as (1), (2), and (3), with the same Burgers vector of $\mathbf{b}=\frac{1}{2}[1\bar{1}0]$.
The characteristic angles between the $\mathbf{b}$ and the line vectors for each dislocation segment are 
$\beta_{(1)}=\ang{19.1}$, 
$\beta_{(2)}=\ang{79.1}$,
and $\beta_{(3)}=\ang{40.9}$.
The closer this characteristic angle is to the value of $\ang{0}$, the more screw character the dislocation segment possesses.
A segment with a characteristic angle nearer to $\ang{90}$ more closely emulates an edge dislocation.
The X-ray beam is incident to the simulation cell in Fig.~\ref{fig:triloop}(a) along the $\mathbf{k}_0$ direction, with the diffraction vector $\mathbf{Q}=[001]$.
The field of view (FOV) sampled by the DFXM image simulations is illustrated as the purple box in Fig.~\ref{fig:triloop}(a).
%
%
Our integrated DD-DFXM algorithm first calculates the deformation gradient $\mathbf{F}^{\rm g}$ induced by the dislocation loop over the voxels defining the illuminated volume (describing the entire image).
The DFXM images can then be computed for each tilt angle $\phi$.
Figure~\ref{fig:triloop}(c) shows the rocking curve for $\phi$ tilt, where the orange area indicates the maximum $I_{\max}$ and minimum $I_{\min}$ intensities (dynamic range), and the black solid line indicates the average $I_{\rm avg}$ intensity.

Within the rocking curve, the tilt $\phi=0$ represents the `strong beam' condition, where the crystal matrix satisfies the Lau\'e condition and the maximal amount of the incoming intensity is diffracted.
In this condition, only the regions near the dislocation lines deviate from the perfect Lau\'e condition -- producing lower image intensities due to the local lattice deformations; this generates the image contrast shown in Fig.~\ref{fig:triloop}(d).
As discussed by~\cite{borgi_simulations_2024}, the strong beam condition is less valid under the geometrical optics.
Therefore, by contrast, a `weak beam' condition may be selected at a tilt of $\phi=\qty{0.4}{\milli\radian}$.
At this condition, the crystal matrix is tilted away from the Lau\'e condition, showing a dark background, with bright dislocation lines, as displayed in Fig.~\ref{fig:triloop}(e).
The rocking curve is asymmetric across the tilt angle $\phi$, as was first discovered and identified as a dislocation effect by~\cite{ungar_x-ray_1984}.
The asymmetry of the rocking curve is also observed in DFXM and discussed in detail in~\cite{pal_measuring_2024}.
%

From the DFXM image in Fig.~\ref{fig:triloop}(d)-(e), the image contrast generated by dislocation segment (1) is much lower than (2) and (3).
As discussed in \cite{pal_measuring_2024}, this indicates that the more the screw component dominates in the Burgers vector $\mathbf{b}=1/2[1\bar{1}0]$ of the dislocation, the lower the contrast in the DFXM images.
The simulation box for this simulation follows the `simplified geometry' defined in~\cite{poulsen_geometrical-optics_2021}, where the crystal system aligns with the Miller's indices, i.e., $\hat{x}_c=[100]$, $\hat{y}_c=[010]$, $\hat{z}_c=[001]$.
With these microscope settings, the local deformed diffraction vectors are calculated by Eq.~(38) in~\cite{poulsen_geometrical-optics_2021},
\begin{equation}
     \mathbf{q}_c = \left[H_{13}+\phi, H_{23} +\chi, H_{33}\right]^T
\end{equation}
where $H_{ij} = \partial u_i/\partial r_j$ is the displacement gradient tensor.
%
%
In the simplified coordinate system, for the screw dislocation ($\hat{\xi}\parallel\mathbf{b}$) with $\mathbf{b}=\frac{1}{2}[1\bar{1}0]$, the displacement only occurs along the line direction $\hat{\xi}$, which has no projection on the $\hat{z}_{\rm g}$ direction. 
Therefore, under the assumption of ideal kinematic diffraction, the diffraction vector $\mathbf{q}_{\rm c} = 0$ for such screw dislocations, thus showing no contrast in the simulated DFXM images.
%


\subsection{Complex dislocation structures from MD simulation}
\label{sec:disl_structures}

\begin{figure}
     \centering
     \includegraphics[width=0.95\linewidth]{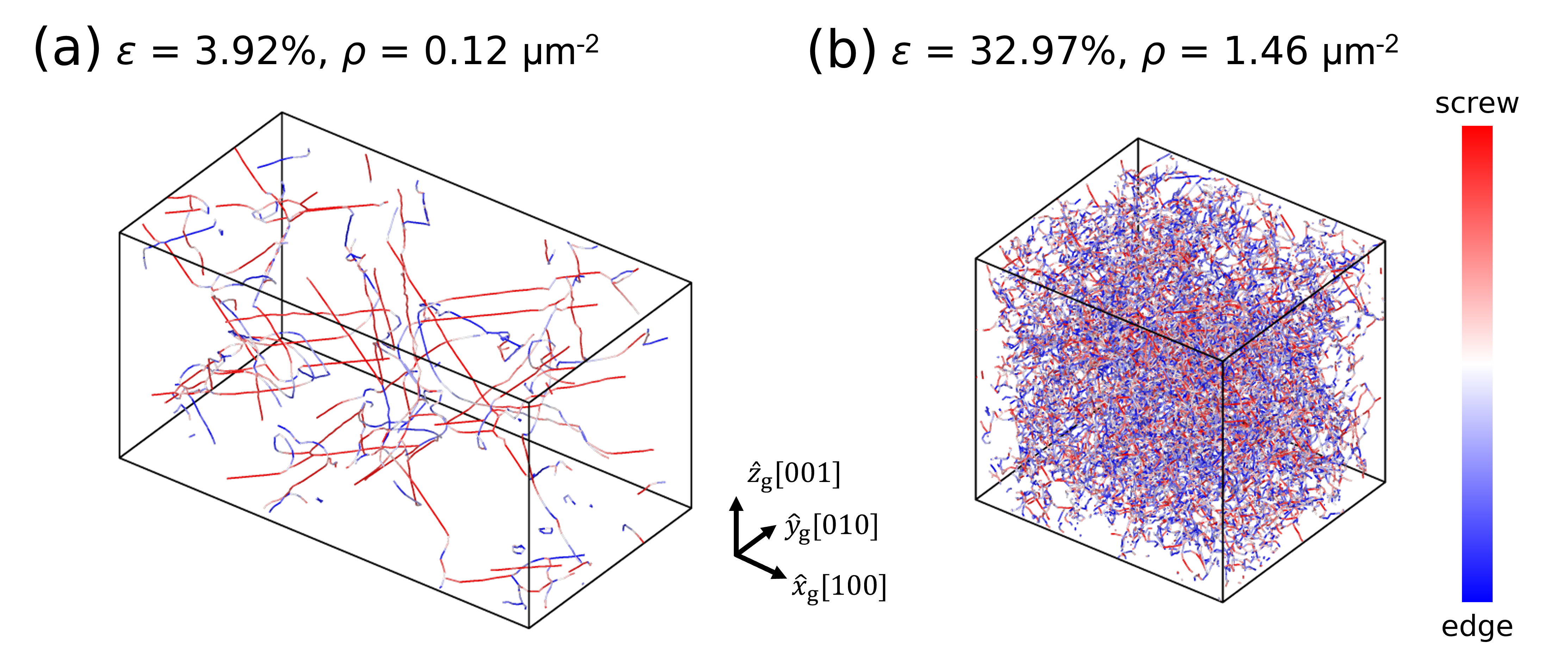}
     \caption{
     {\bf Dislocation structures at different loading stages.}
     %
     %
     Dislocations are shown as linear segments, with colors to indicate the dislocation characteristics based on their Burgers vector.
     The two configurations show (a) the low dislocation density at $\varepsilon = \qty{3.92}{\percent}$, and 
     (b) the high dislocation density configuration at $\varepsilon = \qty{32.97}{\percent}$. Both are under compressive loading.
     }
     \label{fig:configurations}
\end{figure}

Now we consider the case of complex dislocation structures from large-scale MD simulations,
such as the ones shown in Fig.~\ref{fig:configurations}(a)(b).
The simulation is carried out using LAMMPS~\cite{thompson_lammps_2022} on a diamond-cubic crystal of initial aspect ratio 2:1:1 with physical dimensions of $\qtyproduct{\sim 140 x 70 x 70}{nm}$,
oriented along the $[100]\times[010]\times[001]$ crystallographic directions in the simulation frame.
The crystal is subjected to full periodic boundary conditions to simulate the bulk dislocation behaviors under an isothermal, high-strain-rate uniaxial compressive loading.
We use the Stillinger-Weber interatomic potential for Silicon \cite{stillinger_computer_1985} to model a diamond-cubic crystal structure.
Twelve rhomboidal-shaped, prismatic dislocation loops of $\qty{35}{nm}$ radius of the `vacancy type' are introduced randomly in the initially perfect crystal.
Each prismatic loop lies on a randomly selected $\{110\}$ plane with Burgers vector of $\frac{1}{2}\langle110\rangle$, with the two pairs of opposite edges of the rhomboid lying on two different $\{111\}$ slip planes of the Burgers vector.
Each loop is introduced by removing one layer of atoms along the planar cut surface of the loop, resulting in a total of $\num{33452184}$ atoms in the simulation box.

The initial configuration is first equilibrated at a temperature of $\qty{300}{K}$ under zero pressure.
A true strain rate of $\qty{2e8}{s^{-1}}$ in the $x$-direction is then applied to the system by uniformly deforming the simulation box with a compressive loading up to $\qty{50}{\percent}$ strain.
The simulation is carried out at $\qty{300}{K}$ using a Nos\'e-Hoover style dynamics and Langevin thermostat.
The atomic configurations of the MD simulation are saved at different stages of the loading, and their corresponding DD structures are extracted using the dislocation analysis (DXA) algorithm~\cite{stukowski_automated_2012}.

%
%
Our MD simulation enables us to generate realistic DD microstructures characteristic of high-rate deformation loading in diamond-cubic crystals which cannot be generated via other methods.
However, the high computational cost of MD limits our simulation to use box sizes of $\sim\qty{100}{nm}$ in length, which is on par with the resolution of the DFXM technique.
As such, it would be impossible to meaningfully analyze virtual DFXM images computed from the original DD structures generated from MD.
To overcome this issue, we take advantage of the `principle of similitude' for dislocation networks~\cite{kuhlmann-wilsdorf_theory_1985}
and rescale the DD configurations obtained from MD to larger boxes with a minimum side length of $\qty{10}{\mu m}$ (while keeping their original aspect ratio) --
the so-rescaled DD structures being indeed self-similar to the original ones, this rescaling operation therefore preserves the physical nature of the DD configurations.
%
%
%
%
%
%
%
To demonstrate our algorithm, we include two configurations: one at the loading strain of $\varepsilon=\qty{3.92}{\percent}$ and the other at $\varepsilon=\qty{37.92}{\percent}$.
As shown in Fig.~\ref{fig:configurations}, these configurations are selected in this work to demonstrate the characteristics of the DFXM images for systems with low vs high dislocation densities.

\begin{figure}
     \centering
     \includegraphics[width=0.95\linewidth]{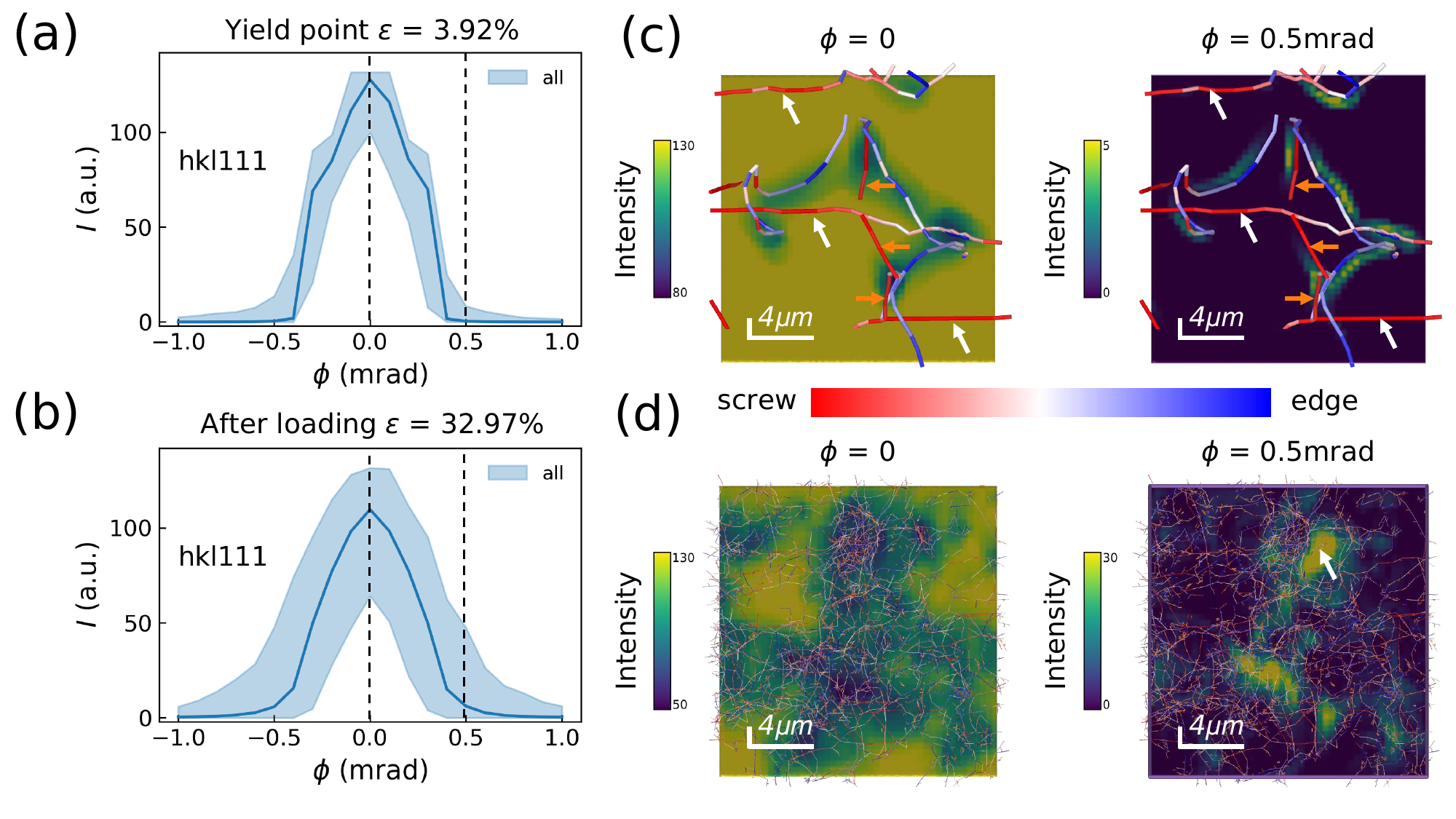}
     \caption{
     {\bf Simulated DFXM images with the $(111)$ diffraction plane.}
     (a) Rocking curves of the low-dislocation-density configuration at the yielding point.
     (b) Rocking curves of the high-dislocation-density configuration after shock compressive loading.
     The light blue area indicates the minimum and maximum intensity as functions of the $\phi$ tilt, and the dark blue curve is the average intensity of the image.
     (c) Simulated DFXM images at the strong beam ($\phi=0$) and weak beam ($\phi=\qty{0.5}{\milli\radian}$) conditions for the low-dislocation-density configuration.
     (d) Simulated DFXM images at the strong beam ($\phi=0$) and weak beam ($\phi=\qty{0.5}{\milli\radian}$) for the high-dislocation-density configuration.
     The intensity map overlays on the DD structures (colored by dislocation character) using the 3D visualization tool OVITO~\cite{stukowski_visualization_2009}.
     Note that the scale bars representing the real space dimensions are not the same on the horizontal and vertical directions, since the real space FOV is projected onto the detector with an angle.
     }
     \label{fig:line_broadening_111}
\end{figure}

The DD-DFXM algorithm is then applied to the DD structures, and the DFXM images are simulated for the $(111)$ diffraction planes, as shown in Fig.~\ref{fig:line_broadening_111}.
In Fig.~\ref{fig:line_broadening_111}(a)(b), we show the mean value of intensity for each configuration's rocking curve scan (dark blue line), overlaid with the full dynamic range of each image at a given $\phi$ tilt position (light blue area), i.e., the intensity range from maximum to minimum.
We note that from these rocking curves, the diffraction peak width broadens as the dislocation density increases, consistent with~\cite{ungar_contrast_1999}.
Beyond this low-dimensional rocking curve analysis, our DFXM simulated images offer a direct view of the source of the broadening and scattering intensities. 
We show this effect in Fig.~\ref{fig:line_broadening_111}(c)(d) for representative images at the strong- ($\phi=0$) and weak-beam ($\phi=\qty{0.5}{\milli\radian}$) conditions.
In addition to the algorithm, we integrate the simulated DFXM workflow with the 3D visualization software OVITO~\cite{stukowski_visualization_2009} widely used by the atomistic simulation community.
The dislocation lines can be overlaid on top of the simulated DFXM images to analyze the dislocation features.
The simulated images in Fig.~\ref{fig:line_broadening_111}(c)(d) are from the detector's point of view.
Thus, the scales in the real coordinate differ for the vertical and horizontal directions, as shown in the scale bar.

The DFXM image for the low dislocation density configuration shows that the contrast comes from each dislocation line.
The edge dislocations are more prominent compared to screw dislocations in strong and weak beam conditions.
The screw dislocation lines show two different types of behavior.
The screw dislocation lines on the diffraction $(111)$ plane show minimum contrast (white arrows), 
while the screw dislocations on the other $\{111\}$ planes (forest dislocations, orange arrows) have contrast and can be observed through DFXM.
%
%
For the high dislocation density configuration (d), the DFXM resolution is insufficient to identify individual dislocations.
The contrast is relatively prominent for locally high-dislocation-density regions in the strong beam condition.
However, we can see a strong contrast in the weak beam image (white arrow) with a relatively low dislocation density.
The contrast mechanism is subject to future research related to detailed dislocation features.

%
%
%

\section{Discussion}

Our work in this study demonstrates a new workflow and the associated computational methods to directly connect dislocation models to continuum-scale DFXM simulations.
This has important implications for macroscopic plasticity behavior, such as strain hardening, creep, twinning, and phase transformation induced plasticity.
%
%
Each of these fields has unique aspects of plasticity that require high-fidelity atomistic or dislocation models to be able to describe in full.
As such, interpretation of the dislocation structures by DFXM has been limited without methods to connect atomistic and DD models to DFXM contrast.

We also note that these approaches offer a unique opportunity to extend the DFXM simulations beyond the dislocation and stacking faults predicted thus far, such as low-angle grain boundaries, twinning boundaries, and shear bands.
%
%
While the DXA method is unique to translating dislocations from MD to DD and enabling the scale connection in this instance, we note that other methods exist that can be more adept for those defect types.

Following the foundational~\cite{wilkens_determination_1970} line profile analysis method, \cite{ungar_x-ray_1984, ungar_dislocation_2001} further found the asymmetry peak broadening due to dislocation structures in single crystal X-ray diffraction~\cite{wilkens_determination_1970, ungar_x-ray_1984, ungar_dislocation_2001},
which can also be observed in our simulated rocking curves.
The current line profile analysis method is based on assumptions of certain dislocation distributions and the corresponding strain fields.
Our DFXM-DDD integrated algorithm can accurately determine the strain field from a dislocation structure, significantly promoting our understanding of the underlying mechanisms of line profiling in single-crystal X-ray diffraction.
Using the model, we can further separate different dislocation characterizations, and analyze their contributions to the contrast in diffraction line profiling.
For a full analysis of the dislocation structures and characters that are resolvable by DFXM, we refer the reader to the sister paper of this work that can be found in~\cite{wang_expanding_2024}. 

\section{Conclusions}
\label{sec:conclusions}

Our integrated DFXM-DDD algorithm advances the interpretation of the imaging for dislocation structures in bulk single-crystal deformation at high dislocation densities relevant to crystal plasticity.
Our tool can also provide unique insights into how the dislocation structures influence line profiling in single-crystal X-ray diffraction analysis.

\appendix

\section{Monte-Carlo ray-tracing algorithm for calculating the resolution function}
\label{sec:ray_tracing}

The reciprocal-space resolution function ${\rm Res}_{\vec{q}_i}(\vec{q}_i)$ describes the variation in values that define the incident wave vector, which dictates the set of possible diffracted wavevectors that could contribute to signal on the detector. The incident wavevector is defined as \ldm{$\vec{k}_{\rm in} = k[\varepsilon, \zeta_h, \zeta_v]$}, where
the energy bandwidth of the incident beam, $\varepsilon = \delta E/E$, the horizontal divergence of the beam, $\zeta_h$, and the vertical divergence of the beam, $\zeta_v$, all define the wavevectors that can give rise to diffraction from the deformed sample, each approximated as Gaussian random variables.
Similarly, the range of diffracted beams $\vec{k}_{\rm out} = k[\varepsilon, \eta, \theta]$, include the wavevector components that may pass through the lens, and is approximated as a truncated Gaussian function based on the numerical aperture of the objective lens that forms the image.
The resolution function can then be calculated by the following equations,
\begin{align}
     q_{\rm rock} &= -\frac{\zeta_v}{2} - (\theta - \theta_0)
     \nonumber \\
     q_{\rm roll} &= -\frac{\zeta_h}{2\sin(\theta)} - \cos(\theta)\eta
     \nonumber \\
     q_{\|} &= \frac{\delta E}{E} + \cot(\theta)\left[-\frac{\zeta_v}{2} + (\theta - \theta_0)\right]
\end{align}
To simplify the calculation, we use the ray-tracing Monte Carlo algorithm to calculate the resolution function, as was performed in~\cite{poulsen_geometrical-optics_2021}.


\section{Einstein's notation}
\label{sec:einstein}

Eq.~(\ref{eqn:dispgrad_seg}) uses Einstein's notation where each subscript $(i,j,k,l,m,p)$ represents a sum over $(1,2,3)$, corresponding to the 3D coordinates $(x,y,z)$, respectively.
For example, $y_i = \sum_{j=1}^3 A_{ij}x_j \equiv A_{ij}x_j$ is equivalent to the matrix multiplication $\mathbf{y}=\mathbf{Ax}$.
The comma operator represents the gradient, for example $F_{,ij} \equiv \frac{\partial^2 F}{\partial r_i\partial r_j}$.
$\mathbf{b} = [b_1, b_2, b_3]^T$ is the Burgers vector;
and $\epsilon_{ijk}=(-1)^p\epsilon_{123}$ is the Levi-Civita symbol, where $p$ is the parity of the permutations, indicating the necessary number of pair permutations to unscramble $i,j,k$ into $1,2,3$.
Here $\epsilon_{123}$ is selected to be $+1$.



\ack{Acknowledgements}

Y.W. was supported by the Stanford Energy Postdoctoral Fellowship.
D.P., K.K., S.J.I., and L.E.D.-M. acknowledge the support from the Air Force Office of Scientific Research (FA9550-23-10347), managed by D. Barbee.
The work of N.B. and R.E.R. was supported by the LDRD program at Lawrence Livermore National Laboratory (21-ERD-032), and performed under the auspices of the U.S. Department of Energy by LLNL under Contract DE-AC52-07NA27344.



\referencelist{iucr}






\end{document}